\documentclass[11pt,final]{scrartcl}

\usepackage{quantum}
\usepackage{paralist}

\usepackage{caption}
\usepackage{graphicx}
\usepackage{epsfig}
\usepackage{tikz}
\usepackage[mode=tex]{standalone}
\tikzset{every picture/.style={line width=0.75pt}} %set default line width to 0.75pt   
\usepackage{bbold}

\usepackage[obeyFinal]{todonotes}

\usepackage{amsmath}
\usepackage{amssymb}
\usepackage{amsthm}

\usepackage{authblk}
\let\email\thanks

\usepackage[normalem]{ulem}

\usepackage[style=alphabetic,doi=false,url=false]{biblatex}
\appto{\bibsetup}{\raggedright}
\bibliography{main}
\renewbibmacro{in:}{}

\newbibmacro{string+doiurl}[1]{%
  \iffieldundef{doi}{%
    \iffieldundef{url}{%
      #1%
    }{%
      \href{\thefield{url}}{#1}%
    }%
  }{%
    \href{http://dx.doi.org/\thefield{doi}}{#1}%
  }%
}

\DeclareFieldFormat{title}{\usebibmacro{string+doiurl}{\mkbibemph{#1}}}
\DeclareFieldFormat[article,incollection]{title}%
    {\usebibmacro{string+doiurl}{\mkbibquote{#1}}}

\definecolor{linkred}{RGB}{160,0,0}
\definecolor{citegreen}{RGB}{0,160,0}
\definecolor{urlblue}{RGB}{0,0,160}
\usepackage[pdfstartview=FitH,      % Oeffnen mit fit width
            bookmarksopen=true,     % aufgeklappte Bookmarks
            bookmarksnumbered=true, % Kapitelnummerierung in bookmarks
            colorlinks=true, urlcolor=urlblue, pdfborder={0 0 0},
            linkcolor=linkred,
            citecolor=citegreen,
            pdfencoding=auto,
            psdextra
            ]{hyperref}
\usepackage[capitalise]{cleveref}

\def \be  {\begin{equation}}
\def \ee  {\end{equation}}
\def \bea {\begin{equation}\begin{aligned}}
\def \eea {\end{aligned}\end{equation}}
\def \ba  {\begin{eqnarray}}
\def \ea  {\end{eqnarray}}

\definecolor{dancolor}{RGB}{181,137,0}
\definecolor{tobycolor}{RGB}{42,161,152}
\definecolor{jlcolor}{RGB}{108,113,196}
\newcommand{\dannote}[2][]{\todo[color=dancolor,#1]{DZ: #2}}

\newcommand{\jlnote}[2][]{\todo[color=jlcolor,#1]{JLB: #2}}
\usepackage{ulem}

\newtheorem{theorem}{Theorem}
\newtheorem{lemma}[theorem]{Lemma}
\newtheorem{algorithm}[theorem]{Algorithm}
\newtheorem{corollary}[theorem]{Corollary}
\newtheorem{proposition}[theorem]{Proposition}

\newcommand{\appropto}{\mathrel{\vcenter{
  \offinterlineskip\halign{\hfil$##$\cr
    \propto\cr\noalign{\kern2pt}\sim\cr\noalign{\kern-2pt}}}}}

\newcommand\ep{\epsilon}

\definecolor{cardinal}{rgb}{0.6,0,0}
\definecolor{darkgreen}{rgb}{0,0.5,0}
\definecolor{golden}{rgb}{0.92, 0.7, 0}
\definecolor{midnight}{rgb}{0, 0, 0.5}
\definecolor{darkblue}{rgb}{0.2, 0, 0.8}

\begin{document}

\title{Dissipative Quantum Gibbs Sampling}

\author[1,2,3]{Daniel Zhang\email{daniel.zhang@phasecraft.io}}
\author[1,4]{Jan Lukas Bosse\email{janlukas@phasecraft.io}}
\author[1,5]{Toby Cubitt\email{toby@phasecraft.io}}

\renewcommand\Affilfont{\small}
\affil[1]{Phasecraft Ltd.}
\affil[2]{St John's College, University of Oxford}
\affil[3]{Mathematical Institute, University of Oxford}
\affil[4]{School of Mathematics, University of Bristol}
\affil[5]{Department of Computer Science, University College London}

\date{\normalsize \today}

\maketitle

\begin{abstract}\vspace{-5ex}
  Systems in thermal equilibrium at non-zero temperature are described by their Gibbs state.
  For classical many-body systems, the Metropolis-Hastings algorithm gives a Markov process with a local update rule that samples from the Gibbs distribution.
  For quantum systems, sampling from the Gibbs state is significantly more challenging.
  Many algorithms have been proposed, but these are more complex than the simple local update rule of classical Metropolis sampling, requiring non-trivial quantum algorithms such as phase estimation as a subroutine.

  Here, we show that a dissipative quantum algorithm with a simple, local update rule is able to sample from the quantum Gibbs state.
  In contrast to the classical case, the quantum Gibbs state is not generated by converging to the fixed point of a Markov process, but by the states generated at the stopping time of a conditionally stopped process.
  This gives a new answer to the long-sought-after quantum analogue of Metropolis sampling.
  Compared to previous quantum Gibbs sampling algorithms, the local update rule of the process has a simple implementation, which may make it more amenable to near-term implementation on suitable quantum hardware.
  This dissipative Gibbs sampler works for arbitrary quantum Hamiltonians, without any assumptions on or knowledge of its properties, and comes with certifiable precision and run-time bounds.
  We also show that the algorithm benefits from some measure of built-in resilience to faults and errors (``fault resilience'').

  Finally, we also demonstrate how the stopping statistics of an ensemble of runs of the dissipative Gibbs sampler can be used to estimate the partition function.
\end{abstract}

\section{Introduction}

Physical systems at thermal equilibrium are expected to be in their Gibbs state.
For quantum mechanical systems, this is the state given by the Boltzmann distribution over its energy eigenstates: $\rho_G = e^{-\beta H}/\cZ$ for a system with Hamiltonian $H$ at inverse-temperature $\beta=1/k_BT$, where $T$ is the temperature, $k_B$ is Boltzmann's constant (conventionally set to $k_B=1$ in natural units), and $\cZ = \tr(e^{-\beta H})$ is the partition function.
Gibbs states therefore play an essential role in thermal physics and statistical mechanics.
Significant research has been devoted to methods for computing or generating Gibbs states on a quantum computer, in particular of many-body Hamiltonians $H=\sum_i h_i$ made up of local (i.e.\ few-body) interactions $h_i$.

Sampling from the Gibbs distribution of a \textit{classical} many-body Hamiltonian up to relative error is contained in the complexity class $\mathsf{BPP}^\mathsf{NP}$.  For quantum systems, there is evidence it is complexity-theoretically even harder~\cite{BravyiGossett}, and is at least $\mathsf{QMA}$-hard. Thus, we do not expect to be able to generate Gibbs states efficiently in general; the run-time must scale exponentially in the number of particles $n$ for certain Hamiltonians.%\footnote{Unless there is a complexity class collapse even more unexpected than $P=NP$.}

For classical statistical mechanics systems, there exists a famous algorithm for sampling from the Gibbs distribution: the Metropolis-Hastings algorithm~\cite{metropolis1953equation,hastings1970monte}.
The Hamiltonian in this case is a real, scalar function $H(x)$ of the (classical) many-body state $x = (x_i)$.
In rough outline, the Metropolis-Hastings algorithm applied to many-body Gibbs distributions consists in starting from some arbitrary initial state, then repeatedly:
\begin{inparaenum}[(i)]
  \item proposing a new state $x'$ from a symmetric proposal
    distribution $q(x' | x)$, e.g.\ by randomly flipping the state
    $x_i$ of a randomly chosen subsystem $i$,
  \item computing $p = e^{-\beta H(x')}/e^{-\beta H(x)}$, and
  \item accepting the new state $x'$ with probability $\min(p,1)$.
\end{inparaenum}
If $H(x)$ is a sum of local terms each depending only on few of the $x_i$, computing $e^{-\beta H(x')} / e^{-\beta H(x)}$ only requires the evaluation of the local terms depending on the flipped subsystem $x_i$. This means that, for such systems, not only the state updates but also the transition probability calculations are local.
This algorithm provably samples from the Gibbs state in the long run: it generates a Markov chain whose fixed point is the Gibbs distribution. Therefore, once the Markov chain has converged, the states $x$ that it generates will be distributed according to $\Pr(x) = e^{-\beta H}/\mathcal{Z}$, as required.
The mixing time of this Markov chain (the time required to converge to the fixed point) is in general exponential in the number of particles $n$ (as it must be given the complexity-theoretic considerations discussed above).
Nonetheless, in practice, the algorithm often converges faster than this and is widely used.

Sampling from \emph{quantum} Gibbs states is harder still.
The key step in the Metropolis-Hastings algorithm involves evaluating the energy $H(x)$ on a state $x$.
For quantum states $\ket{\psi}$, this means measuring the energy $\braket{\psi|H|\psi}$ given only a single copy of $\ket{\psi}$.
This can be done, but it requires quantum phase estimation which is already a non-trivial quantum algorithm acting on the entire quantum state, not just locally on the $i^{\text{th}}$ particle.
Moreover, the algorithm needs to probabilistically either accept the proposed new state $\ket{\psi'}$ or revert to the previous state $\ket{\psi}$.
But measuring the energy using phase estimation collapses the state to an energy eigenstate, so it is not obvious how to recover the state $\ket{\psi'}$ afterwards, nor how to revert to $\ket{\psi}$.
These obstacles were overcome in Ref.~\cite{temme2011quantum} using a ``rewinding trick'' originally developed in a different context~\cite{marriott2005quantum}, to give the first quantum generalisation of the Metropolis-Hastings algorithm.

However, this came at a price: whereas the classical Metropolis-Hastings algorithm is a simple Markov chain with local update rules, the quantum Metropolis algorithm of~\cite{temme2011quantum} requires running a complex quantum circuit, likely requiring a large-scale, fault-tolerant quantum computer before it can be implemented in practice.

Many subsequent works have sought to improve on the original quantum Metropolis algorithm and construct something closer to the simple, local Markov chain of the classical Metropolis algorithm.
The pros and cons of these approaches mirror the corresponding approaches to ground state preparation (summarised in some detail in Ref.~\cite{cubitt_2023}.) See also \cite{chen2023edi} for a recent comparison of existing quantum Gibbs sampling algorithms.

Some of the oldest proposals involve emulating on a quantum computer a model of the physical thermalisation process, by repeatedly coupling the system to a ``thermal bath'' of ancilla qubits, allowing these to interact, and discarding the ancillas~\cite{terhal2000problem, shabani2016artificial,metcalf2020engineered}.
In suitable limits of weak system-bath interactions or (equivalently) small time-steps, Davies~\cite{davies1974markovian,davies1976markovian} showed that these dynamics are described by a Lindblad master equation -- called a Davies generator -- that thermalises the system in the large-time limit.
Implementing this dynamics on a quantum computer either requires implementing time-evolution under the large system-bath Hamiltonian via Hamiltonian simulation algorithms, as well as a large supply of fresh ancilla qubits~\cite{terhal2000problem}. Or it requires implementing the Davies generator directly on the system, which requires quantum phase estimation~\cite{rall2022thermal} or related quantum algorithms such as the operator Fourier transform~\cite{chen2023edi}.

Using Grover's algorithm or related quantum algorithmic techniques can give a quadratic speedup over the direct implementation~\cite{Poulin_2009, chiang2010quantum}.
Further polynomial speedups are possible using modern quantum algorithm approaches~\cite{bilgin2010preparing, yung2012quantum, ozols2013quantum, chowdhury2016quantum, wocjan2021szegedy}.
Analogous to the classical case, adapting classical coupling-from-the-past~\cite{propp1996exact} allows quantum Metropolis methods to sample from the exact Gibbs distribution, rather than an approximation~\cite{francca2017perfect}.
Nonetheless, all of these algorithms require implementing large, global quantum circuits across the whole system, rather than the simple, local updates of classical Metropolis sampling.

Methods based on Quantum/Probabilistic Imaginary Time Evolution (QITE and PITE) involve finding and implementing a large (in general, $\poly\log$ in the system size) quantum circuit implementing a unitary approximation to the Trotterized imaginary-time evolution~\cite{motta2020determining, tan2020fast, silva2021fragmented}.

Variational approaches to Gibbs state preparation~\cite{warren2022adaptive, lee2022variational, getelina2023adaptive, consiglio2023variational}, as well as variational imaginary-time evolution~\cite{mcardle2019variational}, quantum Boltzmann machines~\cite{zoufal2021variational} and Quantum Approximate Optimisation Algorithm (QAOA)-based approaches~\cite{wu2019variational, zhu2020generation} have similar limitations to the Variation Quantum Eigensolver (VQE) ground-state algorithm: they can work well in practice for some systems, but they involve a computationally non-trivial classical optimisation over quantum circuits.
They are moreover heuristic approaches that have no guarantee of outputting the correct state.

Given the practical infeasibility of many of these algorithms on near-term quantum hardware, some effort has been devoted to reducing the resource requirements for specific classes of Hamiltonians or Hamiltonians obeying certain assumptions~\cite{cohn2020minimal, shtanko2021algorithms}, and estimating observables at thermal equilibrium~\cite{lu2021algorithms}.

A different section of the literature studies thermalisation and Davies generators from the rigorous mathematical physics perspective, proving rapid convergence to the Gibbs state under certain conditions, such as commuting Hamiltonians~\cite{kastoryano2016quantum, bardet2023rapid}.
These results are closely connected with non-commutative generalisations of log-Sobolev inequalities~\cite{capel2020modified, bardet2021modified}.

Recently, one of us developed a quantum algorithm, based on a local quantum Markov process constructed from the local terms of the Hamiltonian, which provably converges to the ground state of any quantum Hamiltonian~\cite{cubitt_2023}.
Specifically, the algorithm (termed the ``dissipative quantum eigensolver'' or DQE), proceeds by repeatedly performing weak measurements of the local terms in the Hamiltonian.
In addition to using weak measurements, the key insight in DQE was not to run the quantum Markov chain until it converges to its fixed point, but rather to select a stopping rule conditioned on the measurement outcomes, such that the \emph{stopped} process generates (a good approximation to) the ground state at the stopping time.
Ref.~\cite{cubitt_2023} shows that, despite the ground state not being generated as a fixed point, the DQE algorithm benefits from many of the same advantages as fixed-point Markov algorithms.

Here, we show that a similar local quantum Markov process can be used to sample from the Gibbs state.
By choosing a suitable probabilistic stopping rule, exactly the same family of algorithms that generates the ground state in~\cite{cubitt_2023} can be adapted to instead produce (a good approximation to) the Gibbs state.
We call this the ``dissipative Gibbs sampler'' (DGS).
As in the ground state case, the key insight that allows us to overcome previous obstacles to constructing a local quantum Markov process that samples from the Gibbs state is that the Gibbs state does not appear as the fixed-point of the process, but rather as the state generated at the stopping time of a conditionally stopped process.
Additionally we prove that the dissipative Gibbs sampler inherently benefits from a limited form of resilience to noise and faulty implementation, without any additional overhead, provided the error rate is less than a threshold.

Finally, we also demonstrate how the stopping statistics of an ensemble of runs of the dissipative Gibbs sampler can be used to obtain an estimate of the partition function of the system, up to multiplicative error. Note that, in order to estimate an observable of the Gibbs state, one would generically need to run the dissipative Gibbs sampler multiple times in any case and hence in practical applications one obtains an estimate of the partition function for free.
\dannote[]{Comment on how this jives with Bravyi/Gosset/Chowdhury/Wocjan who show that the QPF problem is poly-time equivalent to the Gibbs sampling problem?}

\section{Dissipative Quantum Gibbs Sampling}\label{sec:DGS}

We first give a general but precise definition of the dissipative Gibbs sampling (DGS) family of algorithms. Throughout, we will use $\norm{\cdot}$ to denote the operator norm and $\norm{\cdot}_1$ the trace norm. We measure distances between quantum channels with the induced 1-norm $\lVert \cE \rVert_1 := \max_{\rho \neq 0} \lVert \cE(\rho) \rVert_1 / \lVert \rho \rVert_1$.
If the Hamiltonian is not clear from context, we write the Gibbs state explicitly as $\rho_G(H) = \frac{e^{-\beta H}}{\cZ}$ and the partition function as $\cZ(H) = \tr(e^{-\beta H})$. Where it is clear from context, we abbreviate these by $\rho_G$ and $\cZ$.

\begin{algorithm}[Dissipative Gibbs Sampler]\label{alg:DGS}
  Let $H = \sum_{i=1}^m h_i$ be a local Hamiltonian, and $\{\cE_0,\cE_1\}$ be the quantum instrument defined by
  \begin{equation} \label{eq:the_instrument_cE}
    \cE_{0}(\rho) = K \rho K, \quad
    \cE_{1}(\rho) = \left(1-\tr\left(K \rho K \right)\right) \rho_0,
  \end{equation}
  where $K$ is a Hermitian operator satisfying $K^2 \leq \1$ and
  $\norm{K - f(H)} \leq \epsilon$ for an injective function $f$.
  Let $0 \leq r_n \leq 1$ for $ n \in \mathbb{N}_0$. The DGS algorithm consists of successively applying the quantum instrument $\{\cE_0,\cE_1\}$ to an initial state $\rho_0$ and, after a run of $n$ zeros, stopping with probability $r_n$ or continuing running with probability $1-r_n$.
\end{algorithm}

We will show that, for suitable choices of parameters, the DGS \cref{alg:DGS} samples from the Gibbs state  $\rho_{G}(H)$.

\begin{theorem}\label{thm:main_theorem}
  \jlnote[inline=true]{Give this theorem a name too? The algorithm and fault resilience have one, so it only feels fair}
  Consider the process of \cref{alg:DGS}, and choose
  \begin{equation} \label{eq:K_product}
        K = \prod_{i=1}^m\Bigl((1-\epsilon)\1 + \epsilon\kappa_i k_i\Bigr)
        \prod_{i=m}^1\Bigl((1-\epsilon)\1 + \epsilon\kappa_i k_i\Bigr)
  \end{equation}
  where
  \begin{equation}
  k_i = \frac{\1-h_i/\norm{h_i}}{2}, \quad
      \kappa = \sum_i\norm{h_i}, \quad
      \kappa_i = \frac{\norm{h_i}}{\kappa}.
  \end{equation}
  Further, choose the probabilities:
  \begin{equation}
    r_n =\frac{ \frac{\lambda^{2n}}{(2n)!}}{\cosh(\lambda)-\sum_{j=0}^{n-1}\frac{\lambda^{2j}}{(2j)!}}
  \end{equation}
  where $\lambda = \frac{\beta \kappa}{\epsilon (1-\epsilon)^{2m-1}}$, and the initial state $\rho_0 = \frac{\1}{D}$. Then the expected state $\E[\rho_{\tau}]$ at the stopping time $\tau$ satisfies:
  \begin{equation}
    \norm{\E[\rho_\tau] -\rho_G}_1 = O(\beta \epsilon \kappa m^2),
  \end{equation}
  and the expected stopping time $\E[\tau]$ is given by:
  \begin{equation}
    \E[\tau] =
    \frac{\cosh(\lambda ) \tr \left(\frac{1}{1-K^2}\right)}{\tr (\cosh(\lambda K)) } -  \frac{ \tr \left( \frac{ K^2 \cosh(\lambda K)}{1-K^{2}} \right)}{ \tr (\cosh(\lambda K)) }
    \leq \frac{6}{\ep} e^{\frac{2\beta \kappa m}{(1-\ep)^{2m-1}}}.
  \end{equation}
\end{theorem}

We emphasise that taking $K$ as in \cref{thm:main_theorem} implies that the quantum instrument has the desirable property of consisting of weak-measuring \emph{local} terms in the Hamiltonian in sequence.
\textcite[Section 9]{cubitt_2023} presents explicit circuit implementations of $K$ in the case where $H$ is a local qubit Hamiltonian.

The run-time bound for the DGS algorithm, in particular the linear scaling with $1/\ep$, is competitive with the recent results of \cite{chen2023edi} which are claimed to be optimal in terms of the scaling with precision.
Furthermore, in the case of DGS, there is no dependence on any (unknown and hard to determine) mixing time, and the algorithm itself has a simple, local implementation.

Note that if we consider the infinite temperature case $\beta = 0$
the parameter $\lambda$ becomes zero and hence $\cosh(\lambda) = 1$ and
$\cosh(\lambda K) = \1$ and the expected run time is
\begin{equation}
  \tau|_{\beta = 0} = \frac{\tr \left[\frac{1}{1-K^{-2}} + \frac{1}{1-K^2}\right]}{\tr \1}
  = \frac{\tr \1}{\tr \1} = 1,
\end{equation}
as expected.

Our strategy to prove \cref{thm:main_theorem} is as follows. In \cref{lem:expected_state} we first derive an expression for the expected state of \cref{alg:DGS} in full generality, before specialising to a specific choice of probabilities $\{r_n\}$ in \cref{cor:cosh_state,lem:explicit_coin_flip_probabilities}. In \cref{lem:ideal_K_cosh_state} we then show that the expected state obtained from running \cref{alg:DGS} with these probabilities and an idealised Kraus operator $K$ is close (in a way we make precise) to the Gibbs state. This idealised $K$, however, is only implementable via global measurements. Combined with two useful \cref{lem:Gibbs_perturbation,lem:K_tilde} on the perturbation of the Gibbs state around a given Hamiltonian, and the closeness of the locally implementable $K$ in \cref{thm:main_theorem} to the idealised one in \cref{lem:ideal_K_cosh_state}, we arrive at the final result. Throughout, we accompany the results on the expected state with corresponding expressions for the expected stopping time of \cref{alg:DGS}.

\begin{lemma}\label{lem:expected_state}
  The expected output state of \cref{alg:DGS} is given by
  \begin{equation}\label{eq:expected_state}
    \E[\rho_\tau] = \frac{\sum_{n=0}^{\infty} r_n R_n \cE_0^{n}(\rho_0)}{\sum_{n=0}^{\infty} r_n R_n \tr \cE_0^{n}(\rho_0)},
  \end{equation}
  where $R_n = \prod_{j=0}^{n-1} (1-r_j)$.
\end{lemma}

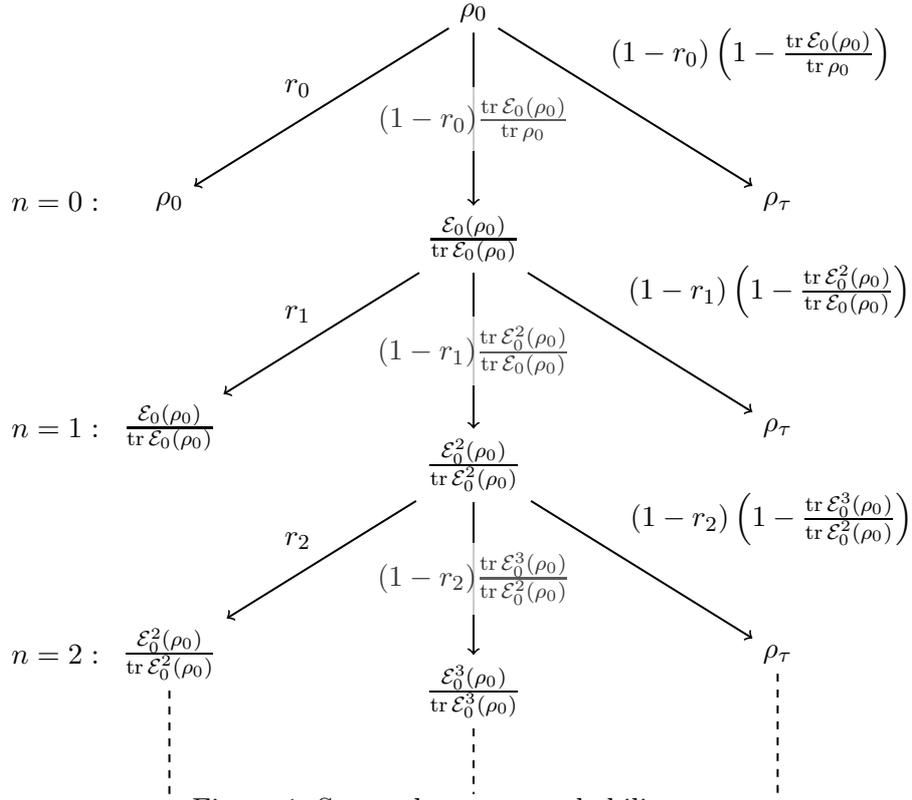
\begin{figure}[h!]
  \centering
  \begin{tikzpicture}
    \node (head) at (0, -0.5) {$\rho_0$};

    \node (n1label) at (-5.5, -3) {$n=0:$};
    \node (n1left) at (-4, -3) {$\rho_0$};
    \node (n1center) at (0, -3.5) {$\frac{\cE_0(\rho_0)}{\tr\cE_0(\rho_0)}$};
    \node (n1right) at (4, -3) {$\rho_\tau$};
    \draw[->] (head) -- (n1left) node[draw=none,midway,above left]{$r_0$};
    \draw[->] (head) -- (n1center) 
      node[draw=none,pos=0.5,fill=white,opacity=0.8,text opacity=1]
      {$(1-r_0)\frac{\tr \cE_0(\rho_0)}{\tr \rho_0}$};
    \draw[->] (head) -- (n1right) node[draw=none,pos=0.4,above right]{$(1-r_0)\left(1- \frac{\tr \cE_0(\rho_0)}{\tr \rho_0}\right)$};

    \node (n2label) at (-5.5, -6) {$n=1:$};
    \node (n2left) at (-4, -6) {$\frac{\cE_0(\rho_0)}{\tr \cE_0(\rho_0)}$};
    \node (n2center) at (0, -6.5) {$\frac{\cE_0^2(\rho_0)}{\tr \cE_0^2(\rho_0)}$};
    \node (n2right) at (4, -6) {$\rho_\tau$};
    \draw[->] (n1center) -- (n2left) node[draw=none,midway,above left]{$r_1$};
    \draw[->] (n1center) -- (n2center) 
      node[draw=none,pos=0.5,fill=white,opacity=0.8,text opacity=1]
      {$(1-r_1)\frac{\tr \cE_0^2(\rho_0)}{\tr \cE_0(\rho_0)}$};
    \draw[->] (n1center) -- (n2right) node[draw=none,pos=0.4,above right]{$(1-r_1)\left(1- \frac{\tr \cE_0^2(\rho_0)}{\tr \cE_0(\rho_0)}\right)$};

    \node (n3label) at (-5.5, -9) {$n=2:$};
    \node (n3left) at (-4, -9) {$\frac{\cE_0^2(\rho_0)}{\tr \cE_0^2(\rho_0)}$};
    \node (n3center) at (0, -9.5) {$\frac{\cE_0^3(\rho_0)}{\tr \cE_0^3(\rho_0)}$};
    \node (n3right) at (4, -9) {$\rho_\tau$};
    \draw[->] (n2center) -- (n3left) node[draw=none,midway,above left]{$r_2$};
    \draw[->] (n2center) -- (n3center) 
      node[draw=none,pos=0.5,fill=white,opacity=0.8,text opacity=1]
      {$(1-r_2)\frac{\tr \cE_0^3(\rho_0)}{\tr \cE_0^2(\rho_0)}$};
  \draw[->] (n2center) -- (n3right) node[draw=none,pos=0.4,above right]{$(1-r_2)\left(1- \frac{\tr \cE_0^3(\rho_0)}{\tr \cE_0^2(\rho_0)}\right)$};

    \node (endleft) at (-4, -11) {};
    \node (endcenter) at (0, -11) {};
    \node (endright) at (4, -11) {};
    \draw[dashed] (n3left) -- (endleft);
    \draw[dashed] (n3center) -- (endcenter);
    \draw[dashed] (n3right) -- (endright);

\end{tikzpicture}
  \caption{Stopped process probability tree}
  \label{fig:probtree}
\end{figure}

\begin{proof}
  We compute the expected state recursively, using the fact that if we measure $1$ at any point, \cref{alg:DGS} dictates we reset to the maximally mixed state $\rho_0$ and begin the process all over again.
  The stopped process may be visualised as the probability tree in \cref{fig:probtree}, where we have replaced each $\rho_0$ we reset to after measuring $1$ with $\rho_\tau$. We have:
  \begin{equation}
    \begin{aligned}
      \E[\rho_\tau] =& \,r_0 \rho_0 + (1-r_0)\left(1-\frac{\tr \cE_0(\rho_0)}{\tr \rho_0}\right)\E[\rho_\tau] \\
      &+ (1-r_0)r_1 \frac{ \cE_0(\rho_0)}{\tr \rho_0}+ (1-r_0)(1-r_1)\left(\frac{\tr \cE_0(\rho_0)}{\tr \rho_0}-\frac{\tr \cE_0^2(\rho_0)}{\tr \rho_0}\right)\E[\rho_\tau]\\
      &+(1-r_0)(1-r_1)r_2 \frac{\cE_0^2(\rho_0)}{\tr \rho_0}+ (1-r_0)(1-r_1)(1-r_2)\left( \frac{\tr \cE_0^2(\rho_0)}{\tr \rho_0} - \frac{\tr \cE_0^3(\rho_0)}{\tr \rho_0}\right)\E[\rho_\tau] \\
      &+ \ldots \\
      &= \sum_{n=0}^\infty r_n R_n  \frac{\cE_0^n(\rho_0)}{\tr \rho_0}
      + \E[\rho_\tau] \sum_{n=0}^\infty R_{n+1} \left( \frac{\tr \cE_0^n(\rho_0)}{\tr \rho_0} - \frac{\tr \cE_0^{n+1}(\rho_0)}{\tr \rho_0}  \right),
    \end{aligned}
  \end{equation}
  where for convenience we define $R_n = \prod_{l=0}^{n-1} (1-r_l)$.
  Noting that
  \begin{equation}
    \begin{aligned}
      \sum_{n=0} &R_{n+1} \left( \tr \cE_0^{n}(\rho_0) - \tr \cE_0^{n+1}(\rho_0) \right)\\
      &= \sum_{n=0} R_n (1 - r_n) \tr \cE_0^{n}(\rho_0) - \sum_{n=0} R_n \tr \cE_0^{n}(\rho_0) + R_0 \tr \rho_0 \\
      &= -\sum_{n=0} r_n R_n \tr \cE_0^{n}(\rho_0) + \tr\rho_0
    \end{aligned}
  \end{equation}
  and $\tr \rho_0 = 1$, we arrive at \cref{eq:expected_state} as claimed.
\end{proof}

The following \namecref{cor:cosh_state} follows immediately.

\begin{corollary}\label{cor:cosh_state}
  If the probabilities $\{r_n\}$ are chosen such that $r_n R_n \propto \frac{\lambda^{2n}}{(2n)!}$ and $\rho_0 = \1/D$, then the expected state produced by \cref{alg:DGS} is
  \begin{equation}
    \E[\rho_\tau] = \frac{\cosh (\lambda K)}{\tr \cosh (\lambda K)}.
  \end{equation}
\end{corollary}
We show that such a choice of $\{r_n\}$ exists in the following \namecref{lem:explicit_coin_flip_probabilities}.

\begin{lemma}\label{lem:explicit_coin_flip_probabilities}
  If we choose the $\{r_n\}$ as
  \begin{equation}
    r_n =\frac{\frac{\lambda^{2n}}{(2n)! }
      }{\cosh(\lambda) -\sum_{j=0}^{n-1}\frac{\lambda^{2j}}{(2j)!}
    }
  \end{equation}
  then we have that $r_n R_n = \frac{\lambda^{2n}}{(2n)! \cosh{\lambda}} $, as required in \cref{cor:cosh_state}.
\end{lemma}

\begin{proof}
  Demanding that $r_n R_n \propto \frac{\lambda^{2n}}{(2n)!}$ is equivalent
  to
  \begin{equation}
    \frac{r_{n+1} \prod_{j=0}^{n} (1-r_j)}{r_n \prod_{j=0}^{n-1} (1-r_j)} = r_{n+1} \frac{1-r_{n}}{r_n} = \frac{\lambda^2}{(2n+2)(2n+1)}.
  \end{equation}
  Straightforward induction shows that this recursive formula for $r_n$ is solved by
  \begin{equation}
    r_n = \frac{r_0 \frac{\lambda^{2n}}{(2n)!}
      }{1-\sum_{j=0}^{n-1} r_0 \frac{\lambda^{2j}}{(2j)!}
    }.
    \label{eq:r_n_solved_recursion}
  \end{equation}

  Additionally, recall that the $\{r_n\}$ must satisfy $0 \leq r_n \leq 1$ for all $n$.
  It is simple to check that this condition is satisfied if and only if $r_0 \leq \frac{1}{\cosh(\lambda)}$.
  We will later see, and it is also intuitively clear, that the expected stopping
  time is minimised if the $\{r_n\}$ are maximal. As the $\{r_n\}$ are increasing with $r_0$, we choose $r_0 = \frac{1}{\cosh(\lambda)}$
  to minimize the run time.
\end{proof}

As an intermediate step, we now show that running \cref{alg:DGS} with an idealised $K$ requiring \emph{global} measurements produces an expected state close to $\rho_G$. We will later show that one can approximate this idealised $K$ sufficiently accurately by weak-measuring \emph{local} terms in the Hamiltonian in sequence.

\begin{lemma} \label{lem:ideal_K_cosh_state}
  If we use $K = (1-\epsilon)^{2m-1}(1 - \frac{\epsilon}{\kappa} H)$
  and $\lambda = \frac{\beta \kappa}{\epsilon (1 - \epsilon)^{2m-1}}$
  in \cref{cor:cosh_state}, the expected output state of \cref{alg:DGS} satisfies
  \begin{equation}
    \Norm{E[\rho_\tau] - \rho_G}_1
     = O(e^{-\frac{\beta \kappa}{\epsilon}}).
  \end{equation}
\end{lemma}

\begin{proof}
  Plugging $K$ and $\lambda$ as above into \cref{cor:cosh_state} allows us to
  calculate
  \begin{equation}
  \begin{aligned}
    \E[\rho_\tau] &= \frac{\cosh(\lambda K)}{\tr \cosh(\lambda K)}
    = \frac{
      e^{\frac{\beta \kappa}{\epsilon} - \beta H} + e^{-\frac{\beta \kappa}{\epsilon} + \beta H}
    }{
      \tr \left(e^{\frac{\beta \kappa}{\epsilon} - \beta H} + e^{-\frac{\beta \kappa}{\epsilon} + \beta H} \right)
    }\\
    &= \frac{\rho_G}{1 + e^{-\frac{2 \beta \kappa}{\epsilon}} \frac{\tr(e^{\beta H})}{\tr e^{-\beta H}}}
    + \frac{
        e^{-\frac{2\beta \kappa}{\epsilon}} e^{\beta H}
      }{
        \tr (e^{-\beta H} + e^{-\frac{2 \beta \kappa}{\epsilon}} e^{\beta H})
      },
  \end{aligned}
  \end{equation}
  which implies
  \begin{equation}
    \norm{E[\rho_\tau] - \rho_G}_1 =
    \Norm{
      \rho_G \left(
        \frac{e^{-\frac{2 \beta \kappa}{\epsilon}} \tr e^{\beta H}}{\tr e^{-\beta H}}
        + o\bigr(e^{-\frac{2\beta \kappa}{\epsilon}}\bigl)
      \right)
      + \frac{
        e^{-\frac{2 \beta \kappa}{\epsilon}} e^{\beta H}
      }{
        \tr \left(e^{-\beta H} + e^{-\frac{2 \beta \kappa}{\epsilon}} e^{\beta H}\right)
      }
    }_1.
  \end{equation}
  The desired result then follows from noting:
  \begin{equation}
    \norm{\rho_G}_1 = 1
    \quad \text{and} \quad
    D e^{- \beta \kappa} \leq \tr e^{\beta H} \leq D e^{\beta \kappa},
  \end{equation}
  and the usual operator norm inequalities.
\end{proof}

We also prove an explicit general expression for the expected stopping time of our algorithm.

\begin{lemma}\label{lem:expected_stopping_time}
  The expected stopping time of \cref{alg:DGS} is
  \begin{equation}\label{eq:expected_stopping_time}
    \E[\tau] = \frac{\sum_{n=0}^{\infty} R_n \tr \cE_0^{n}(\rho_0)}{\sum_{n=0}^{\infty} r_n R_n \tr \cE_0^{n}(\rho_0)}
    \end{equation}
\end{lemma}

\begin{proof}
  This proceeds similarly to the derivation of the expected state.
  Consider again \cref{fig:probtree}, but replace each leaf on the $n\textsuperscript{th}$ level on the left with $n+1$ (the number of steps from the root) and the leaf on the right with $\E[\tau] + n + 1$.
  The expected run time is then given by summing over the values of all leaves weighted by the probability of the path to reach them:
  \begin{equation}
      \E[\tau] = \sum_{n=0} (n+1) r_n R_n  \frac{\tr \cE_0^n(\rho_0)}{\tr \rho_0}
      + (\E[\tau] + n+1) R_{n+1} \left( \frac{\tr \cE_0^n(\rho_0)}{\tr \rho_0} - \frac{\tr \cE_0^{n+1}(\rho_0)}{\tr \rho_0} \right)
  \end{equation}
  and thus
  \begin{equation}
      \E[\tau] = \frac{
        \sum_{n=0} (n+1) r_n R_n \tr  \cE_0^n(\rho_0) + (n+1) R_{n+1} (\tr \cE_0^n(\rho_0) - \tr \cE_0^{n+1}(\rho_0))
      }{
        \tr \rho_0 - \sum_{n=0} R_{n+1} ( \tr \cE_0^n(\rho_0) - \tr  \cE_0^{n+1}(\rho_0))
      }.
    \label{eq:expected_runtime}
  \end{equation}
  The denominator may be simplified identically as in Lemma \ref{lem:expected_state}, and the numerator by using $R_{n+1} = (1-r_n)R_n$ and shifting the summation index of the last term:
  \begin{equation}
    \begin{aligned}
      \sum_{n=0}  (n+1) r_n R_n \tr \cE_0^{n}(\rho_0) + & (n+1) R_n (1-r_n) \tr \cE_0^{n}(\rho_0) - n R_n \tr \cE_0^{n}(\rho_0) \\
      = & \sum_{n=0} R_n \tr \cE_0^{n}(\rho_0).
    \end{aligned}
  \end{equation}
  Note that \cref{eq:expected_runtime} is indeed well defined---i.e.\ the numerator and denominator are both finite---for $\cE_0$
  corresponding to the operator $K$ in both \cref{thm:main_theorem,lem:ideal_K_cosh_state}. This is because, by construction, the
  eigenvalues of $K$ are $\in (0,1)$ for any Hamiltonian with more than one term (i.e.\ $m>1$). Thus we can bound $\tr K^{2n} \leq D
  \lambda_{\text{max}}^{2n}$, and hence if $\lambda_{\text{max}} < 1$ then both
  numerator and denominator converge as $r_n R_n$ and $R_n$ are bounded by $1$.
\end{proof}

To prove \cref{thm:main_theorem} we will also need the following two lemmas from~\cite{Poulin_2009} and~\cite{cubitt_2023}.
They will enable us to show that implementing the quantum instrument in \cref{thm:main_theorem} consisting of weak measurements of local terms in the Hamiltonian produces a state close (in a sense which we make precise) to $\rho_G$.

\begin{lemma}[{\cite[Appendix C]{Poulin_2009}}] \label{lem:Gibbs_perturbation}
  Let $H, H'$ be two Hamiltonians such that $\norm{H-H'} \leq \epsilon$.
  Then
  \begin{equation}
      F(\rho_G(H'), \rho_G(H)) \geq e^{-\beta \epsilon}.
  \end{equation}
\end{lemma}
A simple consequence of this is that, using standard bounds of the trace norm in terms of the fidelity, we may bound
  \begin{equation}\label{eq:Gibbs_perturbation_tracedist}
    \norm{\rho_G(H) - \rho_G(H')}_1 \leq \sqrt{1 - e^{-\beta \epsilon}} = O(\beta \epsilon)
  \end{equation}
if $\norm{H' - H} \leq \epsilon$.

\begin{lemma}[{\cite[Proposition 12]{cubitt_2023}}] \label{lem:K_tilde}
  Let
  \begin{equation}
    \tilde K = (1 - \epsilon)^{2m-1} (1 - \tfrac{\epsilon}{\kappa} H)
  \end{equation}
  and
  \begin{equation}
    K = \prod_{i=1}^m\Bigl((1-\epsilon)\1 + \epsilon\kappa_i k_i\Bigr)
    \prod_{i=m}^1\Bigl((1-\epsilon)\1 + \epsilon\kappa_i k_i\Bigr)
  \end{equation}
  with $\kappa_i$ and $k_i$ as in \cref{thm:main_theorem}. Then
  \begin{equation}
    \norm{\tilde K - K} = O(\epsilon^2 (1 - \epsilon)^{2m-2} m^2)
  \end{equation}
\end{lemma}

We are now in a position to prove the main theorem.

\begin{proof}[Proof of \cref{thm:main_theorem}]
  Let $\tilde{K}$ and $K$ be as in \cref{lem:K_tilde}; this is also the same $K$ as in \cref{thm:main_theorem}. We use \cref{lem:K_tilde} to implicitly define
  $Q$ via $K - \tilde K =  \epsilon^2 (1-\epsilon)^{2m-2} m^2 Q$ with
  $\norm{Q} = O(1)$ and hence write
  \begin{equation}
    H' = H + \frac{\epsilon \kappa m^2}{1-\epsilon} Q
    \quad \text{and} \quad
    K = (1 - \epsilon)^{2m-1}(1 - \tfrac{\epsilon}{\kappa} H').
    \label{eq:H_prime}
  \end{equation}

  \Cref{lem:ideal_K_cosh_state} asserts that running \cref{alg:DGS} with $\cE_0(\rho) = K \rho K^\dagger$ and the $\{r_n\}$ from
  \cref{lem:explicit_coin_flip_probabilities} produces an expected state $\E[\rho_\tau]$ satisfying
  \begin{equation}
     \Norm{\E[\rho_\tau] - \rho_G(H')}_1 = O(e^{-\frac{\beta \kappa}{\epsilon}}).
     \label{eq:difference_gibbs_state_alg_state}
  \end{equation}
  Furthermore, \cref{eq:Gibbs_perturbation_tracedist}
  together with \cref{eq:H_prime} imply  that
  \begin{equation}
    \norm{\rho_G(H') - \rho_G(H)}_1 = O(\beta \epsilon \kappa m^2),
  \end{equation}
  which combined with \cref{eq:difference_gibbs_state_alg_state} and the triangle inequality finally yields
  \begin{equation}
    \norm{\E[\rho_\tau] - \rho_G(H)}_1 = O(\beta \epsilon \kappa m^2) + O(e^{-\frac{\beta \kappa}{\epsilon}}).
  \end{equation}
  This proves the first part of \cref{thm:main_theorem}.

  To compute the expected stopping time, recall from \cref{lem:expected_stopping_time} that
  \begin{equation}\label{eq:expected_stopping_time_proof}
    \E[\tau] = \frac{\sum_{n=0}^{\infty} R_n \tr K^{2n}}{\sum_{n=0}^{\infty} r_n R_n \tr K^{2n}},
  \end{equation}
  and from \cref{lem:explicit_coin_flip_probabilities} that, using the $\{r_n\}$ from \cref{cor:cosh_state},
  \begin{equation}
    \sum_n r_n R_n K^{2n} = \frac{\cosh(\lambda K)}{\cosh(\lambda)}
    \label{eq:unnormalised_state_is_cosh}.
  \end{equation}
  Using $R_{n+1} = R_n - r_n R_n$ we find that
  \begin{equation}
    \sum_{n=0}^{\infty} R_n K^{2n} = K^{-2} \sum_{n=0}^{\infty} R_{n+1} K^{2n+2} + \sum_{n=0}^{\infty}r_n R_n K^{2n},
  \end{equation}
  so that
  \begin{equation}
      (1 - K^{-2}) \sum_{n=0}^{\infty} R_n K^{2n} = \sum_{n=0}^{\infty}r_n R_n K^{2n}  - K^{-2}.
  \end{equation}
  Thus
  \begin{equation}
       \sum_{n=0}^{\infty} R_n K^{2n}
    = \frac{\sum_{n=0}^{\infty} r_n R_n K^{2n} - K^{-2}}{1 - K^{-2}}.
    \label{eq:rewriting_sum_Rn}
  \end{equation}
  Evaluating \cref{eq:expected_stopping_time_proof} by substituting in \cref{eq:unnormalised_state_is_cosh,eq:rewriting_sum_Rn}, we arrive at
  \begin{equation}
    \E[\tau] = \frac{
      \tr \left(\frac{\cosh(\lambda K) - \cosh(\lambda) K^{-2}}{1 - K^{-2}}\right)
    }{
      \tr (\cosh(\lambda K))
    },
  \end{equation}
  demonstrating the exact run-time claimed in the second part of \cref{thm:main_theorem}.
\end{proof}

Finally, we can also upper-bound the expected run time in terms of the parameters of the system, which gives the run-time bound claimed in \cref{thm:main_theorem}.

\begin{proposition}
  The expected run time $\E[\tau]$ of \cref{alg:DGS} for the choice of $K$, $\lambda$, $r_n$ and $\rho_0$ of \cref{thm:main_theorem} is bounded by
  \begin{equation}
    \E[\tau] \leq \frac{\frac{\cosh \lambda}{\cosh (\lambda(1-\ep)^{2m})} }{1- \left(1-\frac{m-1}{m}\ep\right)^{4m}} - \frac{(1-\ep)^{4m}}{1-(1-\ep)^{4m}} := \tau_{\text{max}}.
  \end{equation}
  Further,
  \begin{equation}
    \tau_{\text{max}} \leq \frac{m \, e^{\frac{2\beta \kappa m}{(1-\ep)^{2m-1}}}\left(1+2e^{-\frac{2\beta \kappa}{\ep}}e^{2\beta \kappa}\right)}{(m-1)\ep}  \leq \frac{6}{\ep} e^{\frac{2\beta \kappa m}{(1-\ep)^{2m-1}}}.
  \end{equation}
\end{proposition}

\begin{proof}
    Using the submultiplicativity of the operator norm,
    \begin{equation}
        \Norm{K} \leq \left[ \prod_{i=1}^m (1-\ep) + \ep \frac{\norm{h_i}}{\kappa}\right]^2 \leq \left[\frac{1}{m}\left( \sum_{i=1}^m (1-\ep) + \ep \frac{\norm{h_i}}{\kappa} \right)\right]^{2m},
    \end{equation}
    where the second inequality follows from the arithmetic-geometric mean inequality. As $\kappa = \sum_i\Norm{h_i}$, it follows that
    \begin{equation} \label{eq:agsp_eigenvalue_bounds}
        (1-\ep)^{2m} \leq K \leq \left(1-\frac{m-1}{m}\ep\right)^{2m},
    \end{equation}
    where the lower bound follows from the super-multiplicativity of minimal eigenvalues
    and $0 \leq k_i$ for all $i$.
    Now rewriting $\E[\tau]$ as
    \begin{equation}
    \begin{aligned}
      \E[\tau]
      &=  \frac{
      \cosh{\lambda} \sum_{n=0}^{\infty} \tr (K^{2n}) - \sum_{n=0}^{\infty} \tr(\cosh (\lambda K) K^{2+2n})
      }{
     \tr{(\cosh(\lambda K))}
     } \\
    & \leq   \frac{
      \cosh{\lambda} \sum_{n=0}^{\infty} D  \left(1-\frac{m-1}{m}\ep\right)^{4mn} - \sum_{n=0}^{\infty} D \cosh (\lambda (1-\ep)^{2m}) (1-\ep)^{4m(n+1)}
      }{
      D\cosh(\lambda (1-\ep)^{2m})
      } \\
      & =  \frac{\frac{\cosh \lambda}{\cosh (\lambda(1-\ep)^{2m})} }{1- \left(1-\frac{m-1}{m}\ep\right)^{4m}} - \frac{(1-\ep)^{4m}}{1-(1-\ep)^{4m}}
    \end{aligned}
    \end{equation}
    gives the desired upper bound. The coarse upper bound follows easily after neglecting the second term.
\end{proof}

\section{Partition Functions \& General States}
\label{sec:additional_results}

We now discuss some additional applications of our algorithm. Namely, we describe how to produce an estimate for the partition function $\mathcal{Z}(H)$, as well as sample from a more general class of density matrices which are functions of $K$ (and thus $H$).

First we note that, in a realistic setting, in order to estimate the properties of the Gibbs state, one would need to perform multiple runs of \cref{alg:DGS} to obtain multiple samples from the Gibbs state. If we keep track of the stopping statistics throughout these runs, the next \namecref{prop:partition_functions} demonstrates that we can use these to obtain an estimate for the partition function.

\begin{proposition} \label{prop:partition_functions}
  We can estimate the partition function $\cZ = \tr e^{-\beta H}$ from the
  stopping statistics of an ensemble of runs of \cref{alg:DGS} with the
  $\{r_n\}$ and $K$ as in \cref{thm:main_theorem},
  up to multiplicative error $O(\beta \epsilon \kappa m^2)$. More precisely
  \begin{equation}
    \left|
      D e^{\beta \kappa(2m-1)}
      \E \left[\frac{\text{\# runs}}{\text{\# state resets}}\right] - \cZ(H)
    \right| = O(\cZ(H) \epsilon \kappa m^2 \beta).
  \end{equation}
  Here, ``\# state resets'' is the number of times the state was reset to maximally mixed state (counting \emph{both} resetting due to obtaining the measurement outcome~1, and starting a new run of the algorithm). ``\# runs'' is the number of times \cref{alg:DGS} was run.
\end{proposition}

The proof of \cref{prop:partition_functions} is structured similarly to that of \cref{thm:main_theorem}. We first consider the idealised $\tilde K$ from
\cref{lem:K_tilde} and show the desired result holds up to
$e^{-\frac{\beta \kappa}{\epsilon}}$. We then extend this result to $K$ from
\cref{thm:main_theorem} by absorbing the difference between $\tilde K$ and $K$
into $H$ and using the following \namecref{lem:partition_function_perturbation}.

\begin{lemma}[{\cite[Appendix B]{Poulin_2009}}] \label{lem:partition_function_perturbation}
  Let $H$, $H'$ be two Hamiltonians such that $\Norm{H - H'} \leq \epsilon$. Then
  \begin{equation}
    e^{-\beta \epsilon} \cZ(H') \leq \cZ(H) \leq e^{\beta \epsilon} \cZ(H').
  \end{equation}
\end{lemma}
Assuming that $\beta \epsilon \ll 1$ (as we do throughout this work)
this can be rewritten as
\begin{equation}
  |\mathcal{Z}(H) - \mathcal{Z}(H')| = O(\mathcal{Z}(H) \beta \epsilon)
  \label{eq:partition_function_perturbation}
\end{equation}
by Taylor expanding $e^{\pm \beta \epsilon}$ to first order in $\beta \epsilon$
in both inequalities.

\begin{proof}[Proof of \cref{prop:partition_functions}]
  Considering the paths ending in leaves on the left-hand side of \cref{fig:probtree}
  one can see that the probability to reset the state and then stop
  after $n$ steps (as opposed to having to reset the state again by getting the wrong
  measurement outcome, i.e.\ measuring~1) is given by
  \begin{equation}
    \Pr(\text{stop at $n$}) = r_n R_n \tr \cE_0^n(\rho_0).
  \end{equation}
  With the $\{r_n\}$ and $\cE_0$ as in \cref{thm:main_theorem} and
  using \cref{lem:explicit_coin_flip_probabilities} this becomes
  \begin{equation}
    \Pr(\text{stop at $n$})
    = \frac{\lambda^{2n}}{(2n)! \cosh(\lambda)} \frac{\tr K^{2n}}{D}.
  \end{equation}
  This means that the total probability that \cref{alg:DGS} produces a
  sample after the state was reset, \emph{before} the state is reset again due to a measurement outcome of $1$ from applying the instrument \cref{eq:the_instrument_cE} is:
  \begin{equation}
  \begin{aligned}
    \Pr(\text{produce a sample}) &= \sum_n \Pr(\text{stop at $n$}) \\
    &= \sum_n \frac{\lambda^{2n}}{(2n)! \cosh(\lambda)} \frac{\tr K^{2n}}{D} \\
    &= \frac{\tr \cosh(\lambda K)}{D \cosh(\lambda)}.
  \end{aligned}
  \end{equation}
  Estimating the probability that we successfully produce a sample after
  resetting the state is trivial:
  \begin{equation}
    \E \left[\frac{\text{\# runs}}{\text{\# state resets}}\right]
    = \Pr(\text{produce a sample})
    = \frac{\tr \cosh(\lambda K)}{D \cosh(\lambda)}.
  \end{equation}
  Expanding the $\cosh(\lambda K)$ in terms of exponentials, remembering that
  $K = (1-\epsilon)^{2m-1}(\1 - \frac{\epsilon}{\kappa} H)$ and $\lambda = \frac{\beta \kappa}{\epsilon (1 - \epsilon)^{2m-1}}$, one can solve for
  $\mathcal{Z} = \tr e^{-\beta H}$:
  \begin{equation}
    \cZ = 2 D \cosh(\lambda) e^{-\frac{\beta \kappa}{\epsilon}}
    \E \left[\frac{\text{\# runs}}{\text{\# state resets}} \right]
     - e^{-2\frac{\beta \kappa}{\epsilon}} \tr e^{\beta H}.
     \label{eq:gibbs_estimation_exact_agsp}
  \end{equation}

  The $K$ from \cref{thm:main_theorem} is not exactly of the desired form
  $K = (1-\epsilon)^{2m-1}(\1 - \frac{\epsilon}{\kappa} H)$, but using
  \cref{lem:K_tilde} we can write
  \begin{equation}
    H' = H + \frac{\epsilon \kappa m^2}{1-\epsilon} Q
    \quad \text{and} \quad
    K = (1 - \epsilon)^{2m-1}(\1 - \tfrac{\epsilon}{\kappa} H')
  \end{equation}
  with $\norm{Q} \leq 1$. \cref{eq:partition_function_perturbation} implies then
  \begin{equation}
    |\mathcal{Z}(H) - \mathcal{Z}(H')| = O(\mathcal{Z}(H)\epsilon \kappa m^2 \beta).
  \end{equation}
  Taking this together with \cref{eq:gibbs_estimation_exact_agsp}, expanding
  the $\cosh(\lambda)$ in \cref{eq:gibbs_estimation_exact_agsp} in terms
  of exponentials
  we obtain:
  \begin{equation}
    \left|
      D e^{\beta \kappa (2m-1)}
      \E \left[\frac{\text{\# runs}}{\text{\# state resets}}\right] - \cZ(H)
    \right| = O(\cZ(H) \epsilon \kappa m^2 \beta)
  \end{equation}
  as claimed.
\end{proof}

Another interesting consequence of our analysis is that by choosing the coin-flip probabilities $\{r_n\}$ appropriately in \cref{lem:expected_state}, the expected state of \cref{alg:DGS} can produce:
\begin{equation}
  \E[\rho_\tau] = \frac{f(K)}{\tr f(K)},
\end{equation}
where $f$ is a function obeying certain conditions which we characterise below.
Thus, if one wants to prepare a state whose density matrix is proportional to a function $g(H)$ of $H$, one needs to choose such an $f$ obeying $f(K) = f(1 - \epsilon H) \appropto g(H)$, where $\appropto$ means proportional, up to $O(\ep)$ terms.
Running \cref{alg:DGS} with the modified probabilities $\{r_n\}$, one can then produce (an approximation to) the desired state.
In the case of the Gibbs state this is equivalent to the statement: $\cosh\big(\frac{\beta}{\epsilon} (1-\epsilon H)\big) \appropto e^{-\beta H}$.

\begin{proposition}
    Let $f(x)$ be an even function with power series $f(x) =
    \sum_{n=0}^{\infty} a_n x^{2n}$. Then there exists a choice of coin-flip
    probabilities $\{r_n\}$ such that $r_n \in [0,1]$ and $r_n R_n \propto a_n$
    for all $n$, if and only if either $a_i \geq 0$ for all $i \in
    \mathbb{N}_0$, or $a_i \leq 0$ for all $i \in \mathbb{N}_0$, and further
    $\sum_{i=0}^{\infty} a_i = A$, where $A$ is a finite constant. Running
    \cref{alg:DGS} with these $\{r_n\}$ produces a state $\E[\rho_\tau] \propto f(K)$.
    Additionally, these probabilities are given by:
    \begin{equation}\label{eq:generalprobs}
        r_n = \frac{ca_n}{1-c\sum_{i=0}^{n-1}a_i},
    \end{equation}
    where $c \leq \frac{1}{A}$ if $a_i \geq 0$ and $c \geq \frac{1}{A}$ if $a_i \leq 0$. Since the $r_n$ are monotonic in $c$, choosing $c = \frac{1}{A}$ minimises the run time.
\end{proposition}
\begin{proof}
    Let us first prove sufficiency. If \cref{eq:generalprobs} holds then
    $r_nR_n = c a_n$ follows easily by induction. Further, if all the $ c a_n
    \geq 0$ and $c A \leq 1$ then this implies $r_n \in [0,1]$ for all $n$.

    We now prove necessity. Solving $r_n R_n = c a_n$ again leads to
    \cref{eq:generalprobs}. Demanding that $r_n \in [0,1]$ for all $n \in \N_0$
    in succession implies that we must have $ca_n \geq 0$ and $c
    \sum_{i=0}^{n-1} a_i \leq 1$ for all $n$. Together, these imply the
    conditions specified.
\end{proof}

\section{Faults and Errors}

\Cref{alg:DGS} not only has a similar implementation to the dissipative quantum
eigensolver, it also inherits some of its fault-resilience. 
As a noise model we assume that instead of perfectly implementing
$\cE_0: \rho \mapsto K \rho K$ we only have access to a noisy or perturbed channel
$\cE_0^\prime$ with noise rate $\lVert \cE_0 - \cE_0^\prime \rVert_1 = \delta$.
We also assume that perfectly replacing the state with the maximally mixed one is possible, so that $\cE_1^\prime:\rho \mapsto (1 - \cE_0'(\rho)) \1 / D$. Note that this error model does not assume that the same error occurs with each application of
$\cE_0$, but only that the distribution over possible errors is the same in
each step. It encompasses external noise maps where $\cE_0^\prime = \cN \circ \cE$
as well as mixed channels $\cE_0^\prime = \sum_i p_i \cE_0^{(i)}$ for some
ensemble $\{p_i \cE_0^{(i)}\}$.

\begin{theorem} \label{thm:fault_resilience}
  Let $\cE_0 : \rho \mapsto K \rho K$ be as in \cref{alg:DGS,thm:main_theorem} and $\cE^\prime_0$
  be a perturbed implementation of that channel s.t. the noise rate is
  $\Vert \cE_0 - \cE^\prime_0\Vert_1 = \delta$ with $\delta < \frac{\epsilon}{\beta \kappa}$. Let $\cE_1^\prime: \rho \mapsto  (1 - \cE_0'(\rho)) \1 / D$ be the corresponding perturbed implementation of $\cE_1^\prime$, i.e. we assume perfectly resampling to the maximally mixed state.
  Then the expected state $\E[\rho_\tau]$ produced by \cref{alg:DGS} and
  \cref{thm:main_theorem} and the expected state $\E[\rho^\prime_\tau]$ produced
  with the perturbed implementation $\{\cE_0^\prime, \cE_1^\prime\}$ satisfy
  \begin{equation}
    \Vert \E[\rho^\prime_\tau] - \E[\rho_\tau]\Vert_1
    = O\left(\frac{\delta \beta \kappa}{\epsilon}
      \min \left\{D, e^{2 \beta \kappa} \right\}
    \right).
  \end{equation}
\end{theorem}

\Cref{thm:fault_resilience} tells us that, as long as the noise rate $\delta$ is smaller than $\frac{\epsilon}{\beta \kappa}$, the output state of \cref{alg:DGS,thm:main_theorem} is $O(\delta)$-close to the output of a noiseless implementation.
The output error is not explicitly dependent on the run-time of the algorithm.
However it does scale with inverse-temperature $\beta$, desired precision $\epsilon$, and the problem parameters $\kappa$, $D$.
Because the run-time of DGS also depends on these parameters, this output error is not strictly-speaking independent of the run-time of the algorithm in all parameter regimes.
It is still a non-trivial bound, and in certain parameter regimes (e.g.\ $e^{\beta\kappa} \gg D$), it does scale favourably compared to the total run-time.

The proof of \cref{thm:fault_resilience} consists of collecting useful bounds on the error of the perturbed state
in \cref{lem:perturbation_inequalities} and then using these together with bounding
techniques already used for the proof of \cref{thm:main_theorem}.

\begin{lemma} \label{lem:perturbation_inequalities}
  Let $\cE$ be a completely positive, trace non-increasing map s.t.
  $0 \leq \mu_{\mathrm{min}} \tr(\rho) \leq \tr(\cE(\rho)) \leq \mu_{\mathrm{max}} \tr(\rho)$ for all
  positive $\rho$  and $\cE^\prime$
  be a perturbed implementation of that channel s.t. that in the induced
  1-norm $\Vert \cE - \cE^\prime \Vert_1 = \delta$.
  Then the following inequalities hold:
  \begin{subequations}
    \begin{equation} \label{eq:perturbed_state_bound}
    \left\lVert
      \sum_{n=0}^\infty \frac{\lambda^{2n}}{(2n)!} \cE^n(\1/D)
      - \sum_{n=0}^\infty \frac{\lambda^{2n}}{(2n)!} {\cE^\prime}^n(\1/D)
    \right\rVert_1
    \leq \frac{\lambda \delta}{2 \sqrt{\mu_{\mathrm{max}} + \delta}} \sinh(\lambda \sqrt{\mu_{\mathrm{max}} + \delta})
    \end{equation}
    \begin{equation} \label{eq:perturbed_trace_bound}
    \left\lvert \tr \left[
      \sum_{n=0}^\infty \frac{\lambda^{2n}}{(2n)!} \cE^n(\1/D)
      - \sum_{n=0}^\infty \frac{\lambda^{2n}}{(2n)!} {\cE^\prime}^n(\1/D)
    \right] \right\rvert
    \leq \frac{\lambda \delta}{2 \sqrt{\mu_{\mathrm{max}} + \delta}} \sinh(\lambda \sqrt{\mu_{\mathrm{max}} + \delta})
    \end{equation}
    \begin{equation} \label{eq:normalisation_bound}
    \tr \left[\sum_{n=0}^{\infty} \frac{\lambda^{2n}}{(2n)!} \cE^n(\1/D)\right]
    \geq \max \left\{ \frac{\cosh(\lambda \sqrt{\mu_{\mathrm{max}}})}{D}, \cosh(\lambda \sqrt{\mu_{\mathrm{min}}}) \right\}
    \end{equation}
  \end{subequations}
\end{lemma}

\begin{proof}
  We first prove \cref{eq:perturbed_state_bound}. \Cref{eq:perturbed_trace_bound}
  is then a simple corollary, and \cref{eq:normalisation_bound} will
  follow from the definition of $\mu_{\mathrm{max}}$ and $\mu_{\mathrm{min}}$.

  First, note that for any square matrices $A$ and $B$ and submultiplicative
  matrix norm $\lVert \cdot \rVert$ we have
  \begin{equation} \label{eq:matrix_inequality}
  \begin{aligned}
    \Vert A^n - B^n \Vert &= \left\lVert \sum_{k=0}^{n-1} A^k (A - B) B^{n-1-k} \right\rVert
    \leq \sum_{k=0}^{n-1} \left\lVert A^k (A - B) B^{n-1-k} \right\rVert \\
   &\leq \sum_{k=0}^{n-1} \left\lVert A \right\rVert^k \left\lVert A - B \right\rVert \left\lVert  B \right\rVert^{n-1-k}
   \leq n \Vert A-B \Vert \max\{\Vert A \Vert, B \Vert \Vert\} ^{n-1}.
  \end{aligned}
  \end{equation}
  Using this and that $\lVert \cE^\prime \rVert_1 \leq \mu_{\mathrm{max}} + \delta$
  by the triangle inequality we can show
  \begin{equation}
  \begin{aligned}
    \left\lVert
      \sum_{n=0}^{\infty} \frac{\lambda^{2n}}{(2n)!} (\cE^n - {\cE^\prime}^n) (\1 / D)
    \right\rVert_1
    &\leq
    \left\lVert
      \sum_{n=0}^{\infty} \frac{\lambda^{2n}}{(2n)!} (\cE^n - {\cE^\prime}^n)
    \right\rVert_1 \\
    & \leq
    \sum_{n=0}^\infty \frac{\lambda^{2n}}{(2n)!}
      n \Vert \cE - \cE^\prime \Vert_1
      \max\{\Vert \cE \Vert, \Vert \cE^\prime \Vert\}^{n-1} \\
    & \leq \frac{\lambda \delta}{2}
    \sum_{n=1}^{\infty} \frac{\lambda^{2n-1}}{(2n-1)!} (\mu_{\mathrm{max}} + \delta)^{n-1}.
  \end{aligned}
  \end{equation}
  Summing the last line then gives \cref{eq:perturbed_state_bound}
  as desired.

  \Cref{eq:perturbed_trace_bound} follows from \cref{eq:perturbed_state_bound}
  and the 1-norm bound $\Vert A \Vert_1 \geq \tr(A)$ that is true for any hermitian $A$.

  To show \cref{eq:normalisation_bound} we can on the one hand
  lower bound $\tr \left[\cE^n(\1/D)\right]$ by $\mu_{\mathrm{max}}^{n}/D$ to obtain
  \begin{equation}
    \sum_{n=0}^{\infty} \frac{\lambda^{2n}}{(2n)!} \tr \left[\cE^n(\1/D)\right]
    \geq \frac{\cosh(\lambda\sqrt{\mu_{\mathrm{max}}})}{D},
  \end{equation}
  or also lower bound it by $\mu_{\mathrm{min}}^{n}$ to get
  \begin{equation}
    \sum_{n=0}^{\infty} \frac{\lambda^{2n}}{(2n)!} \tr \left[\cE^n(\1/D)\right]
    \geq \cosh(\lambda \sqrt{\mu_{\mathrm{min}}}).
  \end{equation}
  Taken together, these last two inequalities imply \cref{eq:normalisation_bound}.
\end{proof}

With the inequalities collected in \cref{lem:perturbation_inequalities}
we can now prove the fault resilience result stated in \cref{thm:fault_resilience}.

\begin{proof}[Proof of \cref{thm:fault_resilience}]
  Let $\cE_0$, $\lambda$ and $\cE_0^\prime$ be as in \cref{alg:DGS,thm:main_theorem,thm:fault_resilience}.
  For notational convenience let
  $\rho = \sum_{n=0}^\infty \frac{\lambda^{2n}}{(2n)!} \cE_0^n(\1/D)$ and
  $\rho = \sum_{n=0}^\infty \frac{\lambda^{2n}}{(2n)!} \cE^{\prime n}_0(\1/D)$
  be the un-normalised expected states produced by the noiseless and noisy DGS algorithm.
  We also denote their difference by $\Delta \rho = \rho - \rho'$ and can now
  bound the difference between the normalised expected output states
  as follows:
  \begin{equation}
  \begin{aligned}
    \Vert \E[\rho^\prime_\tau] - \E[\rho_\tau]\Vert_1
     &= \left\lVert \frac{\rho^\prime + \Delta \rho}{\tr(\rho^\prime + \Delta \rho)}
         - \frac{\rho^\prime}{\tr \rho^\prime} \right\rVert_1
     = \left\lVert \frac{\Delta \rho \tr \rho^\prime - \rho^\prime \tr \Delta \rho}{\tr \rho^\prime \tr (\rho^\prime + \Delta \rho)} \right\rVert_1 \\
     & \leq \frac{\Vert \Delta \rho \Vert_1}{\tr (\rho^\prime + \Delta \rho)}
     + \frac{|\tr \Delta \rho|}{\tr(\rho^\prime + \Delta \rho)} \frac{\Vert \rho^\prime \Vert_1}{\tr \rho^\prime} \\
     &= \frac{\Vert \Delta  \rho \Vert_1 + |\tr{\Delta \rho}|}{\tr(\rho^\prime + \Delta \rho)}
     = \frac{\Vert \Delta  \rho \Vert_1 + |\tr{\Delta \rho}|}{\tr\rho}
  \end{aligned}
  \end{equation}
  We now invoke \cref{lem:perturbation_inequalities} with $\cE = \cE_0$ and
  $\cE^\prime = \cE_0^\prime$ to upper bound $\Vert \Delta \rho \Vert_1$
  using \cref{eq:perturbed_state_bound}, $|\tr \Delta \rho|$ with \cref{eq:perturbed_trace_bound} and lower bound $\tr \rho$ by \cref{eq:normalisation_bound}
  to obtain
  \begin{equation}
    \Vert \E[\rho^\prime_\tau] - \E[\rho_\tau]\Vert_1
    \leq \frac{\lambda \delta}{\sqrt{\mu_{\mathrm{max}} + \delta}}
    \min \left\{
      \frac{D \sinh(\lambda \sqrt{\mu_{\mathrm{max}} + \delta})}{\cosh(\lambda \sqrt{\mu_{\mathrm{max}}})},
      \frac{\sinh(\lambda \sqrt{\mu_{\mathrm{max}} + \delta})}{\cosh(\lambda \sqrt{\mu_{\mathrm{min}}})}
    \right\}
    .
  \end{equation}
  In \cref{eq:agsp_eigenvalue_bounds} we showed that
  $(1-\epsilon)^{2m} \leq K \leq (1 - \frac{m-1}{m} \epsilon)^{2m}$ which
  implies for the map $\cE_0: \rho \mapsto K \rho K$ that
  $(1-\epsilon)^{2m} \leq \sqrt{\mu_{\mathrm{min}}}$ and
  $\sqrt{\mu_{\mathrm{max}}} \leq (1 - \frac{m-1}{m} \epsilon)^{2m}$
  and hence $\sqrt{\mu_{\mathrm{max}}} - \sqrt{\mu_{\mathrm{min}}} \leq 2 \epsilon$. Using this and keeping
  only the dominant terms in $\sinh$, $\cosh$ and $\sqrt{\mu_{\mathrm{max}} + \delta}$
  as well as $\delta < \lambda^{-1}$ we obtain the desired result:
  \begin{equation}
  \begin{aligned}
    \Vert \E[\rho^\prime_\tau] - \E[\rho_\tau]\Vert_1
    &= O\left( \lambda \delta
    \min \left\{ D e^{\lambda \delta}, e^{2 \lambda \epsilon} \right\} \right) \\
    &= O\left(\frac{\delta \beta \kappa}{\epsilon}
      \min \left\{D, e^{2 \beta \kappa} \right\}
    \right).
  \end{aligned}
  \end{equation}
\end{proof}

\section{Conclusions}

In this work, we have described an algorithm which provably samples from the
Gibbs state of a generic local Hamiltonian, with certifiable expected run-time
and accuracy. It has numerous desirable properties over previous Gibbs sampling
methods, such as not requiring any promise on the Hamiltonian itself, nor any choice of
ansatz or costly parameter optimisation, as well as only requiring measurements of
local terms in the Hamiltonian.

The DGS algorithm is an adaption of the DQE algorithm from~\cite{cubitt_2023}
to the problem of Gibbs sampling, and therefore shares many of its desirable
properties. For example, in the case of lattice Hamiltonians with finite range
interactions, the Hamiltonian can be split into $O(1)$ sub-Hamiltonians that
only consist of terms acting on disjoint sets of qubits. All terms in such a
sub-Hamiltonian can then be measured in parallel, reducing the time needed for
a single step of the DGS algorithm from $O(m)$ to $O(1)$. Additionally, as in
the DQE algorithm, the DGS algorithm is particularly well-suited to
hardware architectures with ``flying'' qubits which can be readily cycled
through the same circuit and have reliable mid-circuit measurements.
See~\cite[Section~9]{cubitt_2023} for more detailed implementation discussion.

\section{Acknowledgements}
This work was funded by and carried out at Phasecraft Ltd.
T.S.C.~and J.L.B~were supported by EPSRC grant EP/S516090/1.
D.Z.~is also supported by a Junior Research Fellowship from St.~John's College, Oxford.

\printbibliography

\end{document}

%%% Local Variables:
%%% mode: latex
%%% TeX-master: t
%%% End: